# Complex magneto-optic waveguide with dielectric photonic crystal

N. N. Dadoenkova[1,2,*], I. S. Panyaev[1], D. G. Sannikov[1], Yu. S. Dadoenkova[1,2,3], I. A. Rozhleys[4], M. Krawczyk[5], I. L. Lyubchanskii[2]

[1]*Ulyanovsk State University, Ulyanovsk, Russian Federation*
[2]*Donetsk Physical and Technical Institute of the NAS of Ukraine, Donetsk, Ukraine*
[3]*Novgorod State University, Veliky Novgorod, Russian Federation*
[4]*National Research Nuclear University MEPhI, Moscow, Russian Federation*
[5]*Faculty of Physics, Adam Mickiewicz University in Poznań, Poznań, Poland*

A B S T R A C T

We theoretically investigate the dispersion and polarization properties of the electromagnetic waves in a multi-layered structure composed of a magneto-optical waveguide on dielectric substrate covered by one-dimensional dielectric photonic crystal. The numerical analysis of such a complex structure shows interesting properties in propagation of electromagnetic waves, particularly, polarization filtration of TE- and TM-modes depending on geometrical parameters of the waveguide and photonic crystal. We consider different regimes of the modes propagation inside such a structure: when guiding modes propagate inside the magnetic film and decay in the photonic crystal and vice versa; when they propagate in both magnetic film and photonic crystal.

## 1. Introduction

The optical waveguides based on the multilayered photonic structures have been in the focus of intense investigations during last decades. The theory of Bragg reflection waveguides has been proposed in [1, 2] in 1976. It has been shown that waveguides utilizing Bragg reflection at the boundaries can support confined and lossless propagating modes in regions of low refractive index. The waveguide propagation of the electromagnetic waves (EMWs) in nonmagnetic multilayered systems of different geometries, including photonic crystals (PCs) [3], has been reported by many authors [4–6]. The mode selection mechanism is realized by using an asymmetric quasi-one-dimensional Bragg reflection waveguide and shown to be effective to achieve high side-mode suppression ratio. [7]. The guided modes of a slab waveguide which consists of a low-index layer sandwiched between two PCs are analyzed theoretically using a ray-optics model in [4, 5]. It has been shown that the guided modes of the waveguide operate inside the overlapped photonic band gaps of the two Bragg reflectors and each guided mode in such a waveguide has two cutoff points, and the dispersion curves of the guided modes are fragmentary and, as a result, the waveguide can be designed to support only the high-order modes instead of the low-order modes. The theoretical investigations of the quarter-wave Bragg reflection waveguide are presented in [6]. An analytical solution to the mode dispersion equation is derived, and it is shown that such a waveguide is polarization degenerate, although the TE- and TM-mode profiles differ significantly as the external Brewster's angle condition in the cladding is approached. The aperiodic Bragg reflection waveguides have been also studied in Ref. [8].

---

*Corresponding author.
 E-mail address: dadoenkova@yahoo.com (N. N. Dadoenkova).

The structures, combining properties of waveguides and PCs are proposed for many applications. For example, they are used in fabrication of an omnidirectional reflector [9], a high efficiency all-optical diode [10], accelerators [11], mechanically tunable aircore filters [12], polarization splitters [13]. As well the similar structures can be used in the nonlinear regimes to optimize a soliton propagation [14], and for phase matching for frequency conversion [15]. Also an asymmetric Bragg reflection waveguide for a single-mode laser is studied in Ref. [4]. The aperiodic Bragg reflection waveguides have been also investigated to demonstrate a dispersion blue-shift [8]. Effects of linear chirp either in thickness or in refractive index of the cladding layers on the propagation characteristics of one-dimensional photonic band gap planar Bragg reflection waveguides are reported in [16].

However, in the aforementioned papers the attention has been paid to nonmagnetic systems only. The magneto-optic (MO) waveguides and PCs are widely used in modern integrated optics [17] and magneto-photonics [18]. Thus, the combined structures based on the MO waveguide and PCs can open new opportunities in possible application. Until recently, to the best of our knowledge, there was no any publication devoted to the MO waveguides based on the PCs. Last year the MO bilayer sandwiched between two dielectric nonmagnetic Bragg mirrors in combination with plasmonic crystal was studied in Ref. [19]. It was shown that the plasmonic pattern allows excitation of the hybrid plasmonic-waveguide modes localized in dielectric Bragg mirrors of the magneto-photonic crystal or waveguide modes inside its microresonator layer. These modes give rise to the additional resonances in the optical spectra of the structure and to the enhancement of the MO effects [19]. In Ref. [20] the studies of the waveguide modes of one-dimensional magnetic PC with in-plane magnetized layers and nonmagnetic PC with a cladding magnetic layer on the top of the structure were reported.

In this paper, we investigate theoretically a complex waveguide structure composed of MO slab on a dielectric substrate and covered by one-dimensional non-magnetic dielectric PC. In contrast to the aforementioned papers, where the PCs were used as the waveguide claddings to affect the EMWs spectra, we focus on the case when the PC acts as a guiding system as well as the MO slab. The paper is organized as follows. In Section 2 we present the description of the system and provide the analytical calculations of the dispersion of the eigenmodes propagating in the aforementioned hybrid waveguide structure using the transfer matrix method. In Section 3 we discuss different guiding regimes of the normal EMWs and show results of the numerical calculations of the dispersion dependencies. In Section 4, the Conclusions, we summarize the obtained results.

## 2. Model and methods

Let us consider the hybrid MO waveguide system consisting of the magnetic yttrium-iron garnet (YIG) film $Y_3Fe_5O_{12}$ of thickness $L_1$ on the thick dielectric silicon oxide $SiO_2$ substrate and covered by a one-dimensional non-magnetic dielectric PC on the base of gadolinium-gallium garnet (GGG) $Gd_3Ga_5O_{12}$ and titanium oxide $TiO_2$ layers. Thus the system has the structure $SiO_2/YIG/(GGG/TiO_2)^N/vacuum$ with the layers oriented parallel to (*xy*)-plane, the *z*-axis is perpendicular to the interfaces, as shown in Fig. 1. The total thickness of the PC, consisting of $N$ bilayers of GGG/TiO$_2$, is $L_2 = Nd$, where $d$ is the PC's period: $d = d_1 + d_2$ ($d_1$ and $d_2$ are the thicknesses of GGG and TiO$_2$ layers, respectively). It is assumed that the hybrid waveguide structure is of a large extent along the *x*- and *y*-directions which is sufficient to neglect the corresponding boundary effects in these directions. We consider the in-plane magnetization of the YIG layer with the magnetization vector **M** directed along the *y*-axis (see Fig. 1).

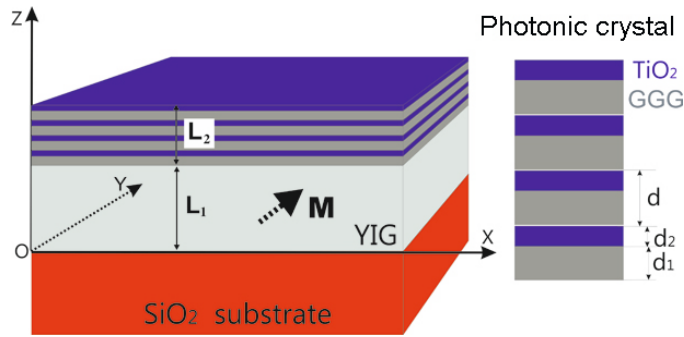

**Fig. 1**. Schematic of the complex waveguide structure $SiO_2/YIG/(GGG/TiO_2)^N/vacuum$. The thicknesses of the magnetic YIG waveguide layer and the covering PC are $L_1$ and $L_2$, respectively, and thicknesses of GGG and TiO$_2$ layers in the PC are $d_1$ and $d_2$, respectively, with $d$ being its period. The thick arrow shows the magnetization direction in YIG layer.

Theoretical analysis of the dispersion of normal EMWs in a bigyrotropic MO waveguide was done in Refs. [21, 22] For obtaining the dispersion equation we use the method described in Ref. [22] in combination with the 4×4 transfer matrix method [23].

We assume that all the layers of the system (YIG, GGG and TiO$_2$) can guide the electromagnetic radiation which propagates along the *x*-axis and exponentially decays in depth of SiO$_2$ substrate and vacuum along the *z*-axis. Propagation of the EMWs in each medium constituting the hybrid waveguide structure is described by the Maxwell's equations [24]:

$$\text{rot}\,\mathbf{E}^{(i)} + \frac{1}{c}\frac{\partial \mathbf{B}^{(i)}}{\partial t} = 0, \quad \text{rot}\,\mathbf{H}^{(i)} - \frac{1}{c}\frac{\partial \mathbf{D}^{(i)}}{\partial t} = 0, \tag{1}$$

with the corresponding material equations for each medium

$$\mathbf{D}^{(i)} = \hat{\varepsilon}^{(i)}\mathbf{E}^{(i)}, \quad \mathbf{B}^{(i)} = \hat{\mu}^{(i)}\mathbf{H}^{(i)}, \tag{2}$$

where $\mathbf{E}^{(i)}$ and $\mathbf{H}^{(i)}$ are the electric and magnetic field vectors, respectively, $\mathbf{D}^{(i)}$ and $\mathbf{B}^{(i)}$ are the electric and magnetic induction vectors, respectively, and the superscripts (*i*) (*i* = 0, 1, 2, 3, m) correspond to vacuum, GGG, TiO$_2$, SiO$_2$, and YIG layers, respectively. In Eq. (2) tensors $\hat{\varepsilon}^{(i)}$ and $\hat{\mu}^{(i)}$ are the dielectric permittivity and magnetic permeability tensors of the corresponding medium.

In the near infra-red regime the cubic YIG is characterized by high transmittivity and exhibits bigyrotropic properties, *i.e.*, both $\hat{\varepsilon}^{(m)}$ and $\hat{\mu}^{(m)}$ tensors contain the off-diagonal components [25]. In the case of **M** || *y* in linear MO approximation the tensors $\hat{\varepsilon}^{(m)}$ and $\hat{\mu}^{(m)}$ of YIG have the following form:

$$\hat{\varepsilon}^{(m)} = \begin{pmatrix} \varepsilon_m & 0 & i\varepsilon' \\ 0 & \varepsilon_m & 0 \\ -i\varepsilon' & 0 & \varepsilon_m \end{pmatrix}, \quad \hat{\mu}^{(m)} = \begin{pmatrix} \mu_m & 0 & i\mu' \\ 0 & \mu_m & 0 \\ -i\mu' & 0 & \mu_m \end{pmatrix}, \tag{3}$$

which corresponds to the transverse MO configuration [26].

The non-magnetic dielectrics GGG, TiO$_2$ and SiO$_2$ are also of cubic symmetry and their dielectric tensors are diagonal: $\hat{\varepsilon}^{(1)} = \varepsilon_1\delta_{ij}$, $\hat{\varepsilon}^{(2)} = \varepsilon_2\delta_{ij}$, and $\hat{\varepsilon}^{(3)} = \varepsilon_3\delta_{ij}$, respectively, where $\delta_{ij}$ is the Kronecker's delta. The magnetic permeabilities of non-magnetic layers are $\mu_i = 1$.

For the layers of the PC and YIG guiding layers, we write the solution $F^{(i)} = \{E^{(i)}, H^{(i)}\}$ of Maxwell's equations Eq. (1) in the form of plane waves:

$$F^{(i)}(x,z,t) = F^{(i)}(z)e^{i(\beta x - \omega t)}, \quad i = 1, 2, m. \tag{4}$$

For vacuum and the SiO$_2$ substrate we write the solution of the Eqs. (1) in the form of plane waves, decaying along the *z*-direction (for $z > L_1 + L_2$) and along the -*z*-direction (for z < 0):

$$\begin{aligned} F_\alpha^{(0)}(z) &= F_\alpha^{(0)}(z)e^{q_0(L_1+L_2-z)}, \\ F_\alpha^{(3)}(z) &= F_\alpha^{(3)}(z)e^{q_0 z}, \end{aligned} \tag{5}$$

where $\alpha = x, y, z$, and $q_{0z}^2 = -\beta^2 + k_0^2\varepsilon_0 < 0$, $q_{3z}^2 = -\beta^2 + k_0^2\varepsilon_3 < 0$ are the *z*-components of the wavevectors in vacuum and SiO$_2$ substrate, respectively. Here $k_0 = \omega/c$ is the modulus of wave vector in vacuum, $\omega$ is the angular frequency of the EMW, and *c* is speed of light in vacuum. In the YIG layer in the transverse MO configuration, the electromagnetic radiation propagating along the *x*-axis splits into independent TE- and TM-polarized EMWs:

$$F_{\text{TE,TM}}^{(m)}(z) = F_{\text{TE,TM}}^{(m)+}(z)e^{ik_{mz}^{\text{TE,TM}}z} + F_{\text{TE,TM}}^{(m)-}(z)e^{-ik_{mz}^{\text{TE,TM}}z}, \tag{6}$$

where $k_{mz}^{\text{TE}}$ and $k_{mz}^{\text{TM}}$ are the wavevector *z*-components of the normal EMWs of TE- and TM-polarizations, respectively:

$$\left(k_{mz}^{\text{TE}}\right)^2 = -\beta^2 + k_0^2\left(n_{\text{YIG}}^{\text{TE}}\right)^2 > 0, \quad \left(k_{mz}^{\text{TM}}\right)^2 = -\beta^2 + k_0^2\left(n_{\text{YIG}}^{\text{TM}}\right)^2 > 0, \tag{7}$$

where we introduce the effective refractive indices of YIG as

$$\left(n_{\text{YIG}}^{\text{TE}}\right)^2 = \varepsilon_m \left(\mu_m^2 - \mu'^2\right)/\mu_m, \qquad \left(n_{\text{YIG}}^{\text{TM}}\right)^2 = \mu_m \left(\varepsilon_m^2 - \varepsilon'^2\right)/\varepsilon_m \qquad (8)$$

Hereinafter $F_{\text{TE,TM}}^{(i)\pm} = \{E_y^{(i)\pm}, H_x^{(i)\pm}\}$ for TE-polarization and $F_{\text{TE,TM}}^{(i)\pm} = \{H_y^{(i)\pm}, E_x^{(i)\pm}\}$ for TM-polarization.

In the GGG and TiO$_2$ layers the electric and magnetic fields of the propagating EMWs have the form similar to that of Eq. (6) with replacing the superscript *m* with *1* for GGG and *2* for TiO$_2$ layers:

$$F_{\text{TE,TM}}^{(1,2)}(z) = F_{\text{TE,TM}}^{(1,2)+}(z)e^{ik_{1,2z}z} + F_{\text{TE,TM}}^{(1,2)-}(z)e^{-ik_{1,2z}z}. \qquad (9)$$

The corresponding *z*-components of wavevectors for GGG and TiO$_2$ are $k_{1z}^2 = -\beta^2 + k_0^2 \varepsilon_1 > 0$ and $k_{2z}^2 = -\beta^2 + k_0^2 \varepsilon_2 > 0$.

At all the interfaces of the hybrid waveguide structure the tangential components of electric and magnetic fields satisfy the boundary conditions:

$$F_{x,y}^{(i)}\Big|_{z=z_j-0} = F_{x,y}^{(l)}\Big|_{z=z_j+0}, \qquad (10)$$

where $i \neq l$, $\{i, l\} = \{1, 2, 3, m\}$, and $z_j$ ($j = 1, 2, 3, ..., (N+1)$) denote the coordinates of the boundaries between all the layers of the system. Solving the system of boundary conditions using 4×4 matrix technique, one can connect the amplitudes of the reflected and transmitted fields:

$$\hat{A}_3 \begin{pmatrix} E_y^{(0)+} \\ H_y^{(0)+} \end{pmatrix} = \hat{T}_g \hat{A}_0 \begin{pmatrix} E_y^{(3)-} \\ H_y^{(3)-} \end{pmatrix}, \qquad (11)$$

where the general transfer matrix $\hat{T}_g = \hat{A}_2 \hat{E}_2(d_2) \hat{S}_{21} \hat{E}_1(d_1) \hat{T}_0^{N-1} \hat{S}_{1m} \hat{E}_m(L_1) \hat{A}_m^{-1}$ with $\hat{T}_0 = \hat{S}_{21} \hat{E}_2(d_2) \hat{S}_{21} \hat{E}_1(d_1)$, being the one period transfer matrix, where the matrices $\hat{S}_{12} = \hat{A}_1^{-1} \hat{A}_2 = \hat{S}_{21}^{-1}$ connect the field amplitudes at the boundaries of the media 1 and 2 (GGG and TiO$_2$ sublayers), and similarly the matrix $\hat{S}_{1m} = \hat{A}_1^{-1} \hat{A}_m$ connects the fields magnitudes at GGG/YIG boundary. The diagonal matrices $\hat{E}_m(L_1)$ and $\hat{E}_j(d_j)$ ($j = 1, 2$), represent the phase incursions within the corresponding layers:

$$\hat{E}_m(L_1) = \text{diag}\left(e^{ik_{mz}^{\text{TE}}L_1}, e^{-ik_{mz}^{\text{TE}}L_1}, e^{ik_{mz}^{\text{TM}}L_1}, e^{-ik_{mz}^{\text{TM}}L_1}\right). \qquad (12a)$$

$$\hat{E}_j(z_j) = \text{diag}\left(e^{ik_{jz}d_j}, e^{-ik_{jz}d_j}, e^{ik_{jz}d_j}, e^{-ik_{jz}d_j}\right), \qquad (12b)$$

The aforementioned matrices $\hat{A}_0$, $\hat{A}_1$, $\hat{A}_2$, $\hat{A}_3$ and $\hat{A}_m$ have the following form:

$$\hat{A}_3 = \begin{pmatrix} \sigma_3^- & 1 \\ 1 & 0 \\ 0 & \gamma_3^- \\ 0 & 1 \end{pmatrix}, \qquad \hat{A}_0 = \begin{pmatrix} \sigma_0^- & 1 \\ 1 & 0 \\ 0 & \gamma_0^- \\ 0 & 1 \end{pmatrix}, \qquad \hat{A}_j = \begin{pmatrix} \sigma_j^+ & \sigma_j^- & 0 & 0 \\ 1 & 1 & 0 & 0 \\ 0 & 0 & \gamma_j^+ & \gamma_j^- \\ 0 & 0 & 1 & 1 \end{pmatrix}, \quad j = 1, 2, m. \qquad (13)$$

with the following coefficients:

$$\gamma_m^\pm = \frac{\pm \varepsilon_m k_{mz}^{\text{TM}} + i\varepsilon'\beta}{k_0(\varepsilon_m^2 - \varepsilon'^2)}, \quad \sigma_m^\pm = \frac{\pm \mu_m k_{mz}^{\text{TE}} + i\mu'\beta}{k_0(\mu_m^2 - \mu'^2)},$$

$$\gamma_i^\pm = \pm\frac{k_{iz}}{k_0 \varepsilon_i}, \quad \sigma_i^\pm = \pm\frac{k_{iz}}{k_0}, \quad (i = 1, 2) \qquad (14)$$

$$\gamma_j^\pm = \pm\frac{q_{3z}}{k_0 \varepsilon_j}, \quad \sigma_j^\pm = \pm\frac{q_{jz}}{k_0}. \quad (j = 0, 3)$$

The requirement of turning to zero the determinant of the system Eq. (11) gives the dispersion relation for the waveguide modes in the considered hybrid structure.

Further we use the analytical results presented above for the numerical calculations of the dispersion dependencies of the normal EMWs in the considered hybrid waveguide structure.

## 3. Numerical results and discussion

### 3.1. Dielectric permittivities and the propagation regimes of the electromagnetic waves

For the numerical calculations we took into account the refractive indices dispersion for all constituent media of the considered hybrid waveguide structure. In the optical and near infra-red regimes the dispersive refractive indices $n_1(\lambda)$ for GGG, $n_2(\lambda)$ for TiO$_2$, and $n_3(\lambda)$ for SiO$_2$, and $n_m(\lambda)$ for YIG have the following forms [27]:

$$n_k^2(\lambda) = \varepsilon_k(\lambda) = f_0^{(k)} + \frac{f_1^{(k)}\lambda^2}{\lambda^2 - \left(\lambda_1^{(k)}\right)^2} + \frac{f_2^{(k)}\lambda^2}{\lambda^2 - \left(\lambda_2^{(k)}\right)^2} + \frac{f_3^{(k)}\lambda^2}{\lambda^2 - \left(\lambda_3^{(k)}\right)^2}, \quad k = 1,2,3,m \quad (15)$$

where the EMW's wavelength $\lambda$ is in microns and the Sellmeier's coefficients $f_\alpha^{(k)}$ and $\lambda_\alpha^{(k)}$ ($\alpha = 0, 1, 2, 3$) are gathered in Table I. In Eq. (15) and the Table 1 the values of $\lambda$ and $\lambda_\alpha^{(k)}$ are in microns.

**Table 1.** Sellmeier's coefficients for the materials constituting the hybrid waveguide structure.

| GGG [27, 28] | $f_0^{(1)} = 1$ | $f_1^{(1)} = 1.7727$ | $f_2^{(1)} = 0.9767$ | $f_3^{(1)} = 4.9668$ |
|---|---|---|---|---|
| | | $\lambda_1^{(1)} = 0.1567$ | $\lambda_2^{(1)} = 0.01375$ | $\lambda_3^{(1)} = 0.22715$ |
| TiO$_2$ [27, 29] | $f_0^{(2)} = 5.913$ | $f_1^{(2)} = 0.2441$ | $f_2^{(2)} = 0$ | $f_3^{(2)} = 0$ |
| | | $\lambda_1^{(2)} = 0.0803$ | $\lambda_2^{(2)} = 0$ | $\lambda_3^{(2)} = 0$ |
| SiO$_2$ [27, 30] | $f_0^{(3)} = 1$ | $f_1^{(3)} = 0.6961663$ | $f_2^{(3)} = 0.4079426$ | $f_3^{(3)} = 0.8974794$ |
| | | $\lambda_1^{(3)} = 0.0684043$ | $\lambda_2^{(3)} = 0.1162414$ | $\lambda_3^{(3)} = 9.896162$ |
| YIG [31] | $f_0^{(m)} = 1$ | $f_1^{(m)} = 3.739$ | $f_2^{(m)} = 0.79$ | $f_3^{(m)} = 0$ |
| | | $\lambda_1^{(m)} = 0.28$ | $\lambda_2^{(m)} = 10.00$ | $\lambda_3^{(m)} = 0$ |

The off-diagonal material tensor elements of YIG are $\varepsilon' = -2.47 \cdot 10^{-4}$ and $\mu' = 8.76 \cdot 10^{-5}$ [21], and for the considered frequency regime $\mu_m = 1$ [25].

In Fig. 2(a) the frequency dependencies of the dielectric permittivities for all the materials constituting the hybrid waveguide structure are shown for the near infra-red regime within the wavelength range from $\lambda_{min} = 1.35$ μm till $\lambda_{max} = 7.77$ μm (which correspond to $\omega_{min} = 2.43 \cdot 10^{14}$ rad·Hz, and $\omega_{max} = 14.00 \cdot 10^{14}$ rad·Hz). The solid (blue), dashed (green), dash-dotted (red) and dotted (black) lines refer to TiO$_2$, YIG, GGG and SiO$_2$ permittivity components, respectively.

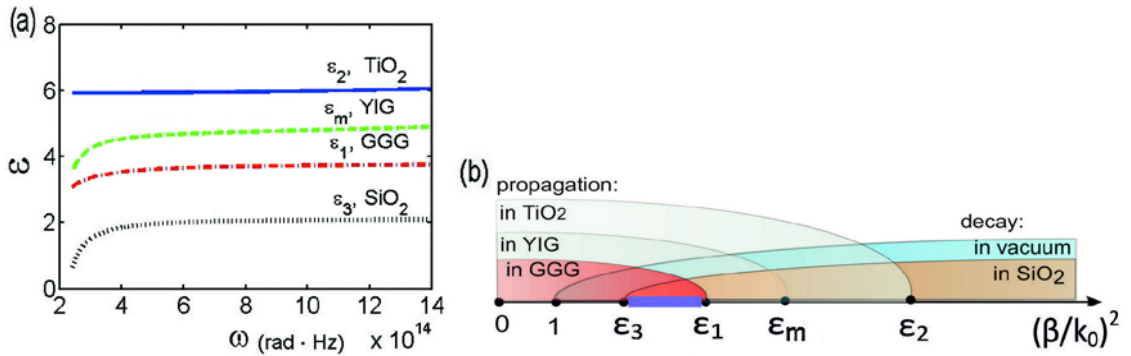

**Fig. 2.** (a) Dielectric permittivities as functions of angular frequency $\omega$ for TiO$_2$ (solid line), YIG (dashed line), SiO$_2$ (dash-dotted line) and GGG (dotted line). (b) Schematic ranges of the EMWs propagation: within the interval $\varepsilon_3 < (\beta/k_0)^2 < \varepsilon_1$ (marked with the dark segment) the EMWs can propagate in YIG layer and all PC's layers; for $\varepsilon_1 < (\beta/k_0)^2 < \varepsilon_m$ the waveguide propagation of EMWs is allowed in YIG and TiO$_2$ layers; for $\varepsilon_m < (\beta/k_0)^2 < \varepsilon_2$ the waveguide propagation is only possible in TiO$_2$ layers.

As one can see from Fig. 2(a), for the considered frequency range, the dielectric permittivity of YIG is larger than that of both the SiO$_2$ substrate and GGG layer, which is the first layer of the PC covering the YIG slab (see Fig. 1). It should be noted that in the considered frequency range the difference between the z-components of wavevectors $k_{mz}^{TE}/k_0$ and $k_{mz}^{TM}/k_0$ in YIG is about 10$^{-9}$, and thus the effective refractive indices introduced in Eqs. (7) and (8) are $n_{YIG}^{TE} \approx n_{YIG}^{TM} = n_m$.

The condition $\varepsilon_3 < \varepsilon_1 < \varepsilon_m$ which is satisfied for the dielectric permittivities of SiO$_2$, GGG and YIG in near infra-red regime, allows an existence of the wavevector interval where the EMWs can propagate simultaneously in YIG layer, as well as in all layers of the dielectric PC, and decay in the substrate and vacuum. In Fig. 2(b) the propagation ranges for the EMWs in YIG, GGG, TiO$_2$ and the ranges of exponential decay of EMWs in vacuum and SiO$_2$-substrate are schematically depicted. The intersection of all these ranges gives the resulting area $\varepsilon_3 < (\beta/k_0)^2 < \varepsilon_1$ which is schematically shown in Fig. 2(b) with the dark segment on the axis $(\beta/k_0)^2$.

Within the interval $\varepsilon_1 < (\beta/k_0)^2 < \varepsilon_m$ the propagation regime changes: the EMWs can propagate in YIG and TiO$_2$ layers as before, but decay in the substrate and GGG layers.

Finally, for $\beta/k_0$ within the interval $\varepsilon_m < (\beta/k_0)^2 < \varepsilon_2$, the EMWs can propagate in TiO$_2$ layers only.

### 3.2. Spectra for the hybrid waveguide structure with the photonic crystal

First we calculate the spectra (reduced wavenumber $\beta/k_0$ vs $\omega$) of the normal EMWs of the hybrid waveguide structures SiO$_2$/YIG/(GGG/TiO$_2$)$^N$/vacuum with increasing the period number N.

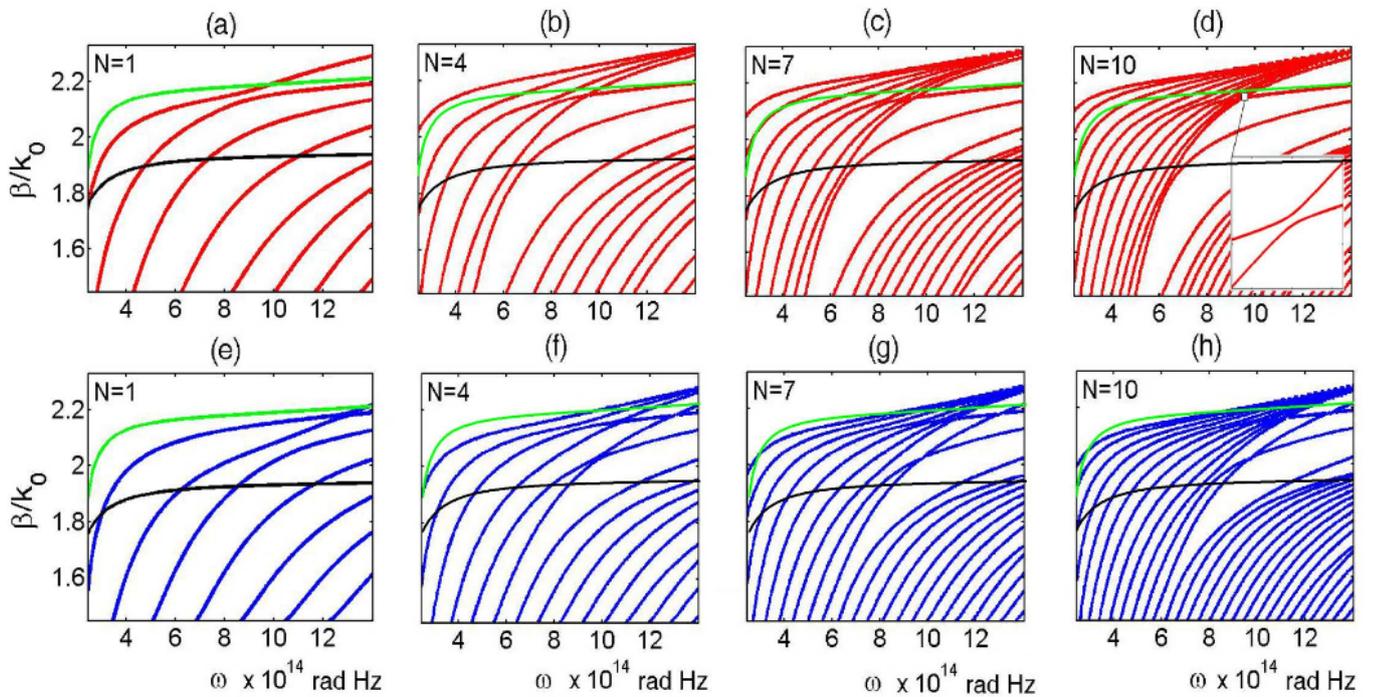

**Fig. 3.** Dispersion curves for the hybrid waveguide structure SiO$_2$/YIG/(GGG/TiO$_2$)$^N$/vacuum: $N = 1$ (a) and (e); $N = 4$ (b) and (f); $N = 7$ (c) and (g); $N = 10$ (d) and (h). The red and blue lines correspond to TE- and TM-polarized modes, respectively. The inset in (d) shows the mini-gap in the vicinity of $\omega \approx 9.5 \cdot 10^{14}$ rad·Hz. The YIG film thickness is $L_1 = 2.0$ μm, the GGG and TiO$_2$ sublayer thicknesses are $d_1 = d_2 = 0.5$ μm. The black and green lines show the refractive index dispersion $n_1(\omega)$ for GGG and $n_{YIG}(\omega)$ for YIG.

In Fig. 3 we show the dispersion curves for the normal EMWs in the case of a single GGG/TiO$_2$ bilayer ($N = 1$) covering the YIG layer for TE- and TM-modes, respectively [Figs. 3(a) and 3(e)]; $N = 4$ [Figs. 3(b) and 3(f)], $N = 7$ [Figs. 3(c) and 3(g)], and for $N = 10$ [Figs. 3(d) and 3(h)]. The red and blue lines correspond to TE- and TM-polarized EMWs, respectively. The calculations are performed for the YIG film thickness $L_1 = 2.0$ μm, and the GGG and TiO$_2$ sublayer thicknesses $d_1 = d_2 = 0.5$ μm. The black and green lines in Figs. 3(a)-3(h) show the refractive index dispersion for GGG $n_1(\omega) = \sqrt{\varepsilon_1(\omega)}$ and YIG $n_{YIG}(\omega)$, respectively, which indicate the ranges of changing the EMWs' propagation regime, discussed above. The dispersion lines (or their parts) located below the black line characterize the EMWs propagating in all layers of the structure except the SiO$_2$ substrate. Within the intermediate region $n_1(\omega) < \beta/k_0 < n_{YIG}(\omega)$ [the range between the black and green lines in Fig. (3)] the propagation of EMWs is possible in YIG and $TiO_2$ layers, while for the range above the green line, the EMWs propagate in $TiO_2$ layers only.

As one can see from Figs. 3(a)-3(e), the increase of $N$ leads to increase of the number of dispersion lines in the considered frequency range as well as to forming some gaps in the spectra. These gaps are similar to the photonic band gaps in the transmittivity spectra of the PCs and they become more pronounced for larger values of $N$. The behavior of the TE- and TM-polarized modes differs for all values of $N$: for TE-modes the gap is formed already at $\beta/k_0 = 1.44$ for $N = 7$, while for TM-modes the gap is forming when $\beta/k_0$ increases the value of about 1.70 for $N \geq 7$. In Figs. 3(b)-3(d), 3(f)-3(h) one can see the family of dispersion curves which has a tendency for asymptotic convergence at the high frequencies. Some dispersion lines approach each other (without intersection) and form mini-gaps. As an example, one of such mini-gaps is shown in the inset to Fig. 3(d), where it appears in the specter of TE-modes at the frequency range $(9.4 \cdot 10^{14} < \omega < 9.6 \cdot 10^{14})$ rad·Hz. The mini-gap width does not exceeds $0.01\ \beta/k_0$ (dimensionless wave number). The similar gap for the TM-mode ($N = 7$) is slightly shifted to the higher frequencies $(9.6 \cdot 10^{14} < \omega < 9.9 \cdot 10^{14})$ rad·Hz. The number of mini-gaps depends on the polarization of EMWs. For example, for $N = 4$ there in only one mini-gap for TE-modes at about $\omega \approx 9.5 \cdot 10^{14}$ rad·Hz [see Fig. 3(b)], while for TM-modes one can see three mini-gaps [Fig. 3(f)] in the vicinity of $\omega \approx 9.8 \cdot 10^{14}$ rad·Hz, $\omega \approx 10.7 \cdot 10^{14}$ rad·Hz and $\omega \approx 13.0 \cdot 10^{14}$ rad·Hz. The number of mini-gaps increases with the increase of $GGG/TiO_2$ bilayer's number $N$ in the system (see Fig. 3).

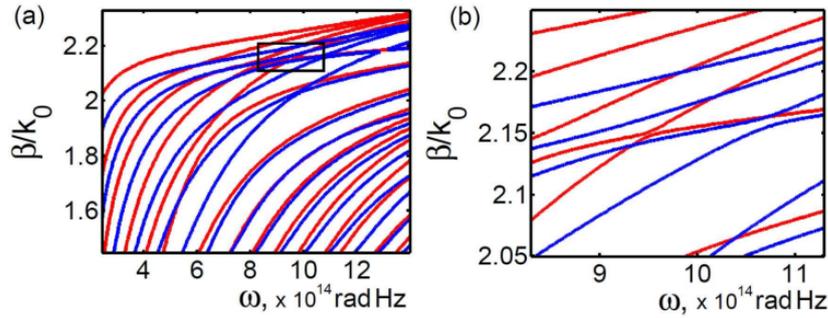

**Fig. 4.** The spectra of normal EMWs of TE- and TM-polarizations (red and blue lines, respectively), for the structure $SiO_2/YIG/(GGG/TiO_2)^N/vacuum$ in the vicinity of the degeneracy points (intersection of dispersion curves for TE- and TM-modes).

It should be mentioned that the spectra of normal EMWs for the structures $SiO_2/YIG/(GGG/TiO_2)^N/vacuum$ demonstrate presence of the degeneracy points, where the dispersion curves of TE- and TM-modes intersect. In Fig. 4(a) we show the spectra for the waveguide structure with $N = 4$ ($GGG/TiO_2$) bilayers with several cross-points (intersection of dispersion curves for TE- and TM-modes).

We also study the influence of the YIG layer thickness $L_1$ on the dispersion curves of the hybrid waveguide structure. In Figs. 5(a)-5(h) we show the evolution of the spectra for TE- (red lines) and TM-modes (blue lines) for $L_1 = 0, 1, 2$, and 4 μm. The calculations are performed for the case of the PC with $N = 7$ periods. Similarly to Fig. 3, the black and green lines in Figs. 5(a)-5(h) indicate the ranges of changing the EMWs' propagation regime. As it was discussed above, within the range between the black and green lines in Figs. 5(b)-5(d) and Figs. 5(f)-5(h), the propagation of EMWs is possible in YIG and $TiO_2$ layers, while for the range above the green line the EMWs propagate in $TiO_2$ layers only. As one can see from Figs. 5(a) and 5(e), in absence of the YIG layer ($L_1 = 0$) (i.e., for the structure $SiO_2/(GGG/TiO_2)^7/vacuum$), the pronounced gaps in the spectra are formed for both TE- and TM-polarized EMWs. The dispersion curves (or the part of dispersion curves) below the black line $n_1(\omega)$ characterize the EMWs, propagating in all the layers forming the photonic structure $(GGG/TiO_2)^N$, and those above the black line correspond to propagation of the EMWs in $TiO_2$ layers only.

Introducing the YIG layer, we obtain the narrowing of the gaps in the EMWs spectra for both polarizations; as well some dispersion curves form the mini-gaps. For example, for TM-polarized mode the mini-gap appears in the vicinity of $\omega \approx 10.8 \cdot 10^{14}$ rad·Hz in the case of YIG layer thickness $L_1 = 1$ μm [see Fig. 5(f)], while for TE-mode [Fig. 5(b)] in the vicinity of $\omega \approx 7.7 \cdot 10^{14}$ rad·Hz the dispersion lines come close one to another but do not intersect.

In Figs. 5(a)-5(h) one can see that the increase of YIG layer thickness leads to increase of the dispersion curves number and the number of mini-gaps within the considered frequency range both for TE- and TM-polarized EMWs.

Introducing the YIG layer, we obtain the narrowing of the gaps in the EMWs spectra for both polarizations; as well some dispersion curves form the mini-gaps. For example, for TM-polarized mode the mini-gap appears in the vicinity of $\omega \approx 10.8 \cdot 10^{14}$ rad·Hz in the case of YIG layer thickness $L_1 = 1$ μm [see Fig. 5(f)], while for TE-mode [Fig. 5(b)] in the vicinity of $\omega \approx 7.7 \cdot 10^{14}$ rad·Hz the dispersion lines come close one to another but do not intersect.

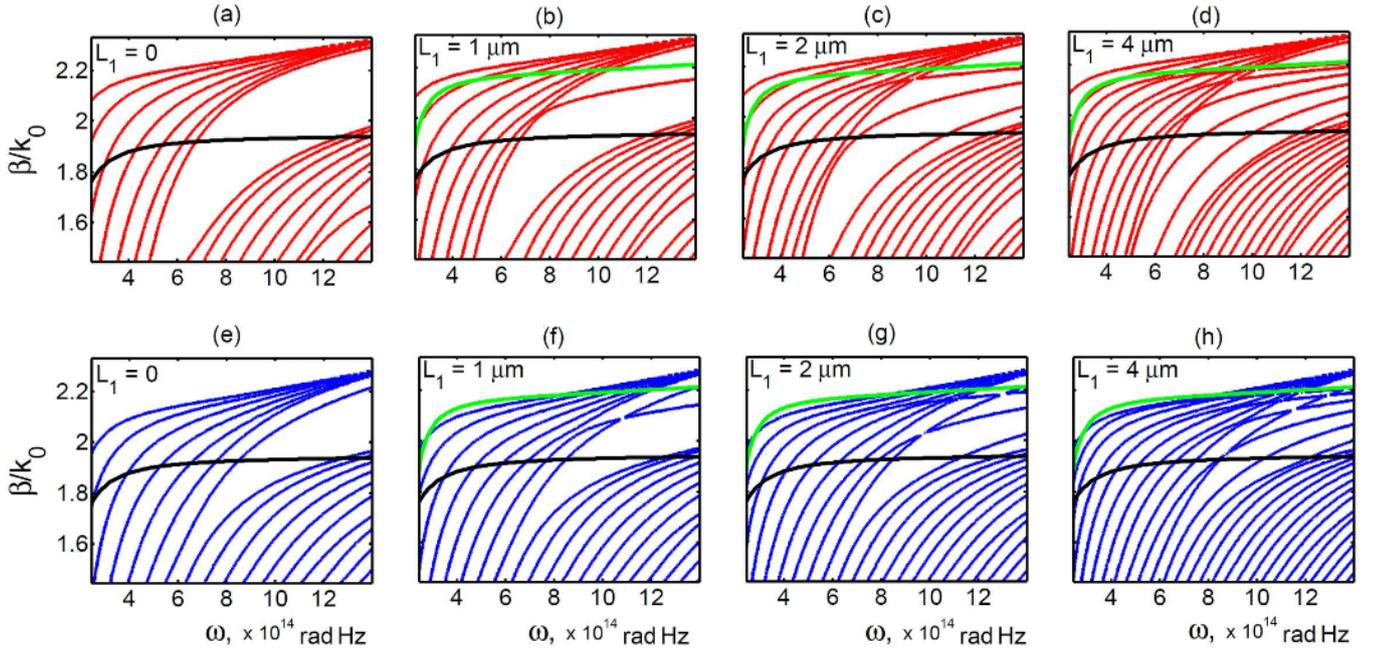

**Fig. 5.** Dispersion curves for the structure $SiO_2/YIG/(GGG/TiO_2)^N/vacuum$ with $N = 7$ bilayers for different thicknesses of the YIG layer: $L_1 = 0$ (a) and (e); $L_1 = 1.0$ μm (b) and (f); $L_1 = 2.0$ μm (c) and (g); $L_1 = 4.0$ μm (d) and (h). The red and blue lines correspond to TE- and TM-polarized modes, respectively. The GGG and $TiO_2$ layer thicknesses are $d_1 = d_2 = 0.5$ μm. The black and green lines show the refractive index dispersion for GGG $n_1(\omega)$ and YIG $n_{YIG}(\omega)$, respectively.

In Figs. 5(a)-5(h) one can see that the increase of YIG layer thickness leads to increase of the dispersion curves number and the number of mini-gaps within the considered frequency range both for TE- and TM-polarized EMWs.

As the dispersion spectra are sensitive to the YIG layer thickness, we can thus fix the EMW wavelength and investigate in details the dependence of the wavenumber $\beta/k_0$ on $L_1$. For the numerical calculations at the working wavelengths $\lambda_0 = 1.55$ μm we used the following geometrical parameters of the waveguide structure: $L_1$ varies from zero to 10 μm, and $d_1 = d_2 = 0.5$ μm. In Fig. 6 we present the dispersion curves (the reduced waveguide propagation number $\beta/k_0$ as function of the reduced YIG layer thickness $L_1/\lambda_0$ for the complex MO waveguide structure $SiO_2/YIG/(GGG/TiO_2)^N/vacuum$ with different numbers $N$ of $GGG/TiO_2$ bilayers. The red and blue lines correspond to TE- and TM-modes, respectively. The black horizontal line defined as $\beta/k_0 = n_1(\lambda_0) = 1.935$ and the green one $\beta/k_0 = n_{YIG}(\lambda_0) = 2.201$ separate the regions, where the EMWs have different propagation regime. The dispersion curves located below the black line correspond to the EMWs propagating in all the layers of the system, except the substrate and vacuum. The dispersion lines between the black and green lines characterize the EMWs propagating in YIG and $TiO_2$ layers only.

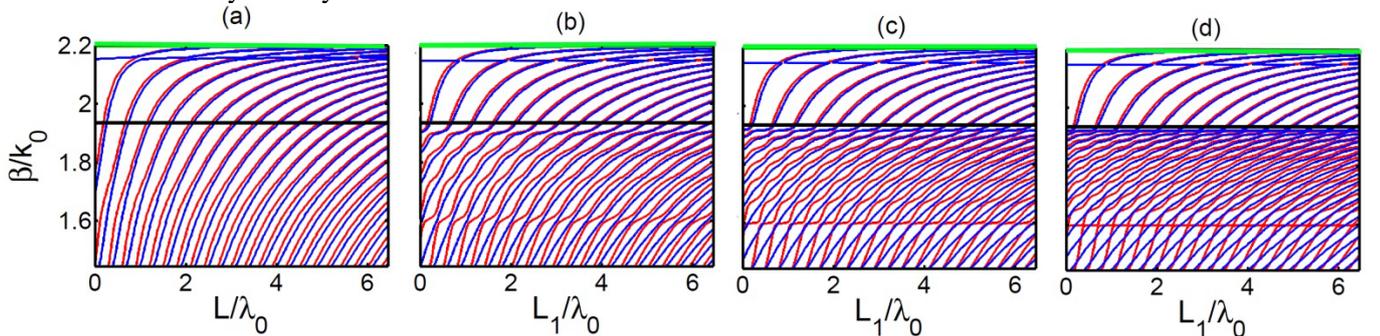

**Fig. 6**. Dispersion curves for the structure $SiO_2/YIG/(GGG/TiO_2)^N/vacuum$: (a) $N = 1$, (b) $N = 4$, and (c) $N = 7$ and (d) $N = 10$. The YIG film thickness changes from 0 to 10 μm, the $TiO_2$ and $SiO_2$ sublayer thicknesses are $d_1 = d_2 = 0.5$ μm. The red and blue lines correspond to TE- and TM-modes, respectively. The calculations are performed for $\lambda_0 = 1.55$ μm. The black horizontal line ($\beta/k_0 = n_1(\lambda_0) = 1.935$) and the green one ($\beta/k_0 = n_{YIG}(\lambda_0) = 2.201$) separate the regions, where the EMWs have different propagation regimes.

The dispersion curves shown in Fig. 6(a) are calculated for the structure with the single $GGG/TiO_2$ bilayer ($N = 1$) covering the YIG waveguide layer. In Figs. 6(b), 6(c) and 6(d) the resulting dispersion curves for the complex waveguides with the dielectric PCs with $N = 4$, $N = 7$ and $N = 10$ are presented, respectively. Comparing

Figs. 6(a)-6(d) one can see that the increase of the number of the PC's periods leads to essential deformation of the dispersion dependencies, namely, to corrugated form of these curves. As well one can see the flattening areas about $\beta/k_0 \approx 1.93$, and $\beta/k_0 \approx 2.2$ where the dispersion curves become almost parallel to the abscissa axis. The upper curves for TE- and TM-modes almost merge in the vicinity of $\beta/k_0 \approx 2.2$. The flattening regions located near $\beta/k_0 \approx 1.93$ become more pronounced with increase of the period number $N$. The increase of N also leads to appearance the degeneracy points, where the dispersion curves of TE- and TM-modes intersect [see Figs. 6(c) and 6(d)] and the modes of both polarizations may coexist. For all values of N the spectra of TM-modes have the mini-gap at about $\beta/k_0 \approx 2.0$, which narrows for big values of $N$.

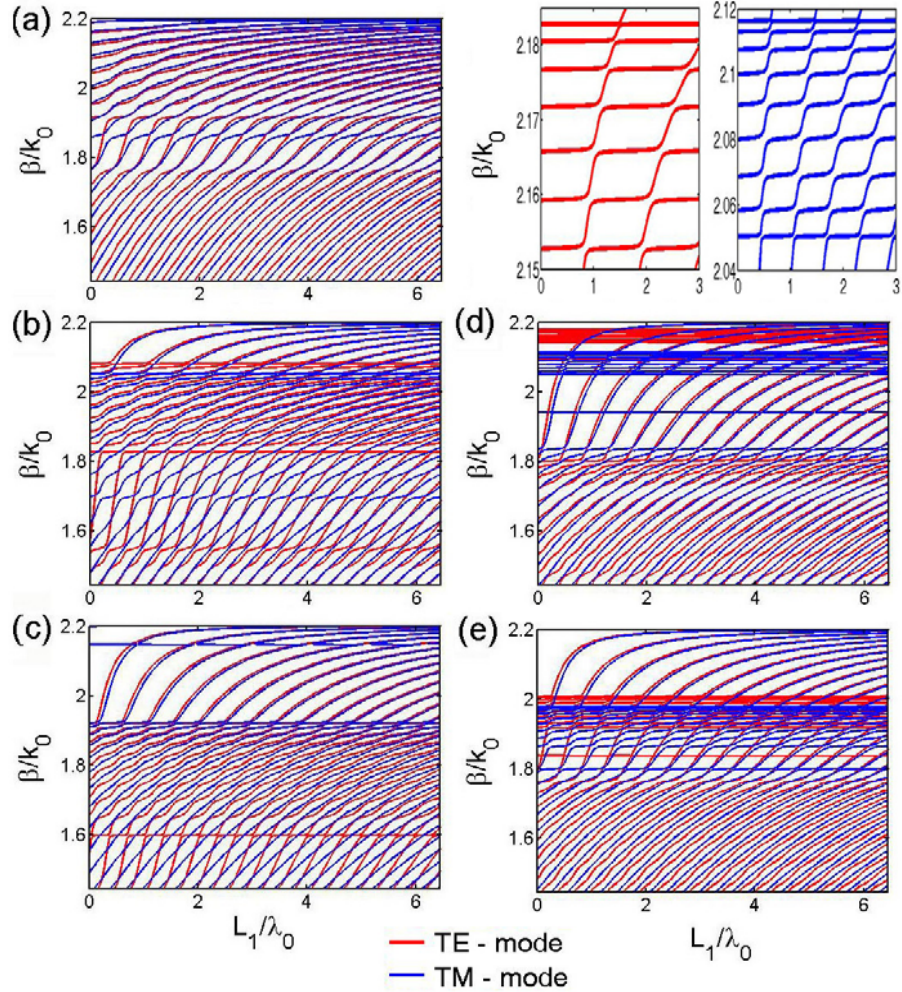

FIG. 7. Evolution of the hybrid waveguide spectra ($\beta/k_0$ vs $L_1/\lambda_0$) with variation of the thicknesses of the PC's layers $d_1$ and $d_2$. The PC period $d = d_1 + d_2 = 1$ μm is constant, and $N = 10$ is fixed: (a) $d_1 = 0.1$ μm $d_2 = 0.9$ μm, (b) $d_1 = 0.3$ μm, $d_2 = 0.7$ μm, (c) $d_1 = 0.5$ μm, $d_2 = 0.5$ μm, (d) $d_1 = 0.7$ μm, $d_2 = 0.3$ μm, (e) $d_1 = 0.9$ μm, $d_2 = 0.1$ μm. The calculations are performed for $\lambda_0 = 1.55$ μm.

Finally, in Fig. 7 we present the evolution of the waveguide spectra ($\beta/k_0$ vs $L_1/\lambda_0$) with variation of the thicknesses of the PC's layers $d_1$ and $d_2$. The PC period is fixed to be the constant value $d = d_1 + d_2 = 1$ μm, while $d_1$ and $d_2$ change: in Fig. 7(a) $d_1 = 0.1$ μm $d_2 = 0.9$ μm, in Fig. 7(b) $d_1 = 0.3$ μm, $d_2 = 0.7$ μm, in Fig. 7(c) $d_1 = 0.5$ μm, $d_2 = 0.5$ μm, in Fig. 7(d) $d_1 = 0.7$ μm, $d_2 = 0.3$ μm, and in Fig. 7(e) $d_1 = 0.9$ μm, $d_2 = 0.1$ μm. For all these calculations we choose the number of the dielectric PC periods $N = 10$, the working wavelength is $\lambda_0 = 1.55$ μm. From comparison of Figs. 7(a)-7(e) it follows that the considered spectra are very sensitive to the relation between the dielectric layer thicknesses. Thus, the increase the GGG layer thickness $d_1$ (with corresponding decrease of the TiO_2 layer thickness $d_2$) leads to appearance of the new horizontal parts in the spectra of TE- and TM modes. For example, for $d_1 = 0.7$ μm, $d_2 = 0.3$ μm [see Fig. 7(d)], we obtain the system of narrow mini-gaps in the regions $2.15 < \beta/k_0 < 2.19$ for TE-modes and $2.04 < \beta/k_0 < 2.12$ for TM-modes, as shown in the insets to Fig. 7(d). The further increase of the GGG rate to $d_1 = 0.9$ μm ($d_2 = 0.1$ μm) in the PC's structure gives the lowering of the positions of these mini-gaps on the $\beta/k_0$ axis to $1.92 < \beta/k_0 < 2.01$ for TE-modes and $1.86 < \beta/k_0 < 1.98$ for TM-modes [Fig. 7(e)].

## 3.3. Limiting cases of the waveguide structure without photonic crystal

In the limiting cases, when the thickness of one of the dielectric sublayer constituting the PC becomes zero, we obtain the four-layer waveguide systems: *SiO$_2$/YIG/TiO$_2$/vacuum* with $d_1 = 0$, and *SiO$_2$/YIG/GGG/vacuum* with $d_2 = 0$. In Figs. 8(a) and 8(b) we show the dispersion curves $\beta/k_0$ vs $\omega$ for *SiO$_2$/YIG/TiO$_2$/vacuum* (a) and *SiO$_2$/YIG/GGG/vacuum* (b) structures. The red and blue lines refer to TE- and TM- polarized modes, respectively.

There are two waveguide regimes in the structure under consideration: the regime A, where there are two guiding layers, and the regime B with the only guiding layer, while the other one plays a role of either a substrate or a cladding [32]. The dash-dotted line, corresponding to $n_{YIG}(\omega)$, and dashed line, referring to $n_1(\omega)$, become in fact the cutoff limits for the double-layer guiding modes (regime A), as shown in Figs. 8(a) and 8(b), respectively. The region between the dashed and dash-dotted lines corresponds to the regime B. For example, Fig. 8(a) shows the TiO$_2$-layer guiding modes (the YIG layer plays a role of a substrate), while the YIG-layer guiding modes (the GGG layer plays a role of a cladding) are represented in Fig. 8(b). The waveguide characteristics of all TE- and TM- polarized modes are similar to ones in Fig. 3, calculated with the matrix method, however, using the effective medium approximation, one can identify a mode order by a number of nodes of the transverse field distributions, or profile function for TE- and for TM-modes. The zeroth fundamental mode has the highest mode index $\beta/k_0$ at any frequency in comparison with the next order mode, *etc*. With the increasing mode order its effective refractive index decreases at a fixed frequency. In general the propagation constants for the TE-modes are larger than for the TM-modes of the same order throughout the whole frequency range. The existence area of the modes is restricted by $n_3$ (SiO$_2$) from below and by $n_2$ (TiO$_2$) [Fig. 8(a)] or by $n_1$ (GGG) [Fig. 8(b)] from above.

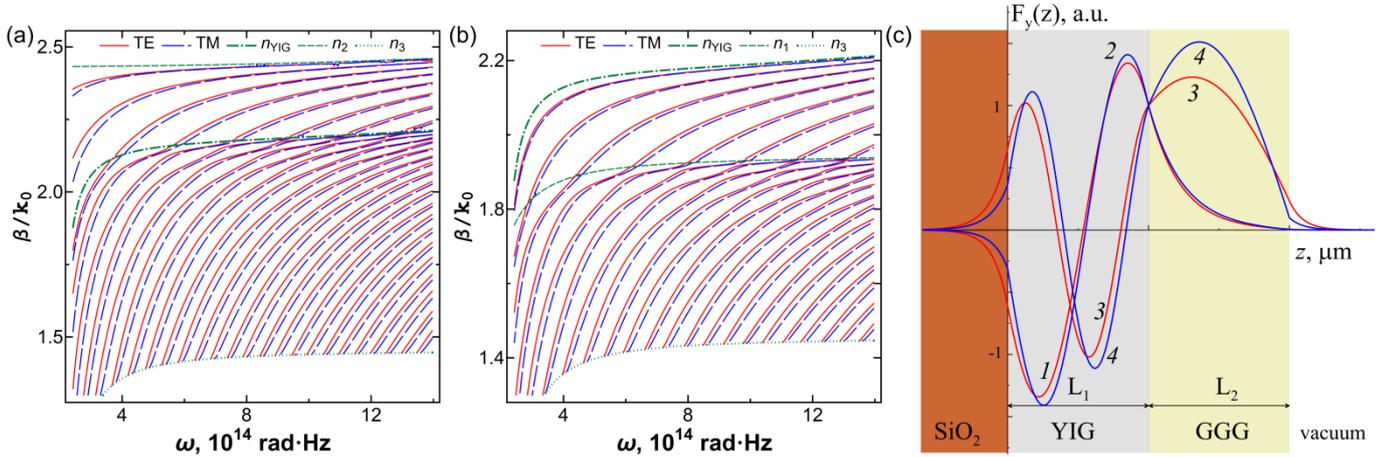

**Fig. 8**. Dispersion curves for *SiO$_2$/YIG/TiO$_2$/vacuum* structure (a), *SiO$_2$/YIG/GGG/vacuum* structure (b) and field profile functions along the *z*-axis in the case of *SiO$_2$/YIG/GGG/vacuum* structure (c) for the TE- and TM-modes (solid red and dashed blue curves, respectively) for $\omega = 4 \cdot 10^{14}$ rad·Hz, $L_1 = 5$ μm, $L_2 = 5$ μm; lines 1, 2 correspond to the TE$_1$- and TM$_1$-modes, and lines 3, 4 correspond to the TE$_2$ and TM$_2$-modes, respectively.

The profile functions of the YIG guiding TE$_1$- and TM$_1$-modes (regime B) are represented in Fig. 8(c) for $\omega = 4 \cdot 10^{14}$ rad·Hz, $L_1 = 5$ μm, $L_2 = 5$ μm by the lines *1* and *2*, which have only one node within the guiding layer. The TE$_2$- and TM$_2$-modes at the same frequency are below the cutoff limits, therefore these modes are the four-layer waveguide modes of the regime A [see lines 3 and 4 in Fig. 8(c)].

## 3.4. Spectra for the hybrid waveguide structure with the effective medium approximation

In the case of thin GGG and TiO$_2$ layers, when $(d_1, d_2) \ll \lambda_0$ (*i.e.*, when we deal with the long wavelength limit for the PC) the effective medium approximation can be applied to present the dielectric PC as a homogeneous anisotropic slab of thickness $L_2$ with the effective dielectric permittivity tensor $\hat{\varepsilon}^{eff}$. In this approach the considered hybrid MO waveguide structure can be treated as an asymmetric four-component waveguide *SiO$_2$/YIG/effective medium/vacuum* which can be described by the method similar to one proposed in Ref. [33]. The components of $\hat{\varepsilon}^{eff}$ are of the form:

$$\varepsilon_{xx}^{eff} = \varepsilon_{yy}^{eff} = \frac{\varepsilon_1 + \varepsilon_2}{1 + d_1/d_2}, \quad \varepsilon_{zz}^{eff} = \frac{(1 + d_1/d_2)\varepsilon_1\varepsilon_2}{\varepsilon_1 + \varepsilon_2}. \tag{16}$$

The normal EMWs of the effective non-magnetic medium are the TE- and TM-modes with the wavevectors

$$\left(k_{eff,z}^{TE}\right)^2 = -\beta^2 + k_0^2 \varepsilon_{yy}^{eff}, \tag{17a}$$

$$\left(k_{eff,z}^{TM}\right)^2 = -\left(\varepsilon_{xx}^{eff}/\varepsilon_{zz}^{eff}\right)\beta^2 + k_0^2 \varepsilon_{xx}^{eff} \tag{17b}$$

Thus, the PC as the effective medium can be characterized by two effective refractive indices $n_{eff}^{TE}$ and $n_{eff}^{TM}$.

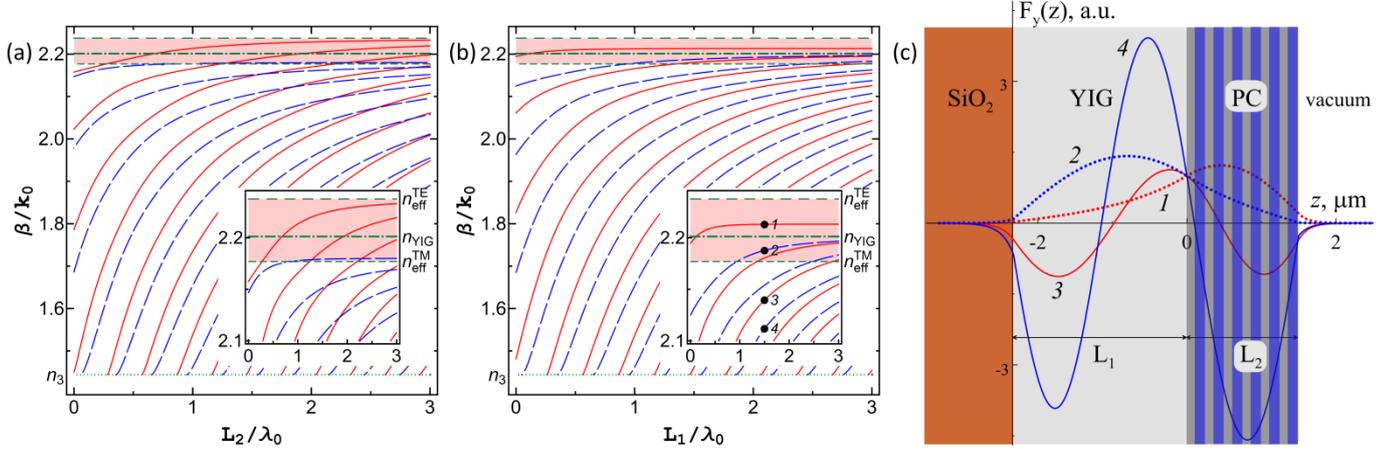

**Fig. 9**. Dispersion spectra $\beta/k_0$ vs $L_2/\lambda_0$ (a) and $\beta/k_0$ vs $L_1/\lambda_0$ (b) of the four-layer structure *SiO$_2$/YIG/effective medium/vacuum* for TE- and TM-modes (solid red and dashed blue curves, respectively) and the asymptotes $n_{YIG}$, $n_{eff}$ (dash-dotted and dashed lines, respectively) for $\lambda_0 = 1.55$ μm, $d_1/d_2 = 0.8$, $L_1 = 1.5$ μm (a) and $L_2 = 1.5$ μm (b). (c) The field profile functions along the z-axis for the TE and TM modes (red and blue lines, respectively) for $L_1/\lambda_0 = 1.5$ (dotted lines 1 and 2: regime B for TE$_0$ and TM$_0$-modes; solid lines 3 and 4: regime A for TE$_2$ and TM$_2$-modes).

In Figs. 9(a) and 9(b) we present the dispersion curves and the field profile functions in the long wavelength limit for the PC. The calculations are carried out for the PC of thickness $L_2 = 1.5$ μm, $d_1/d_2 = 0.8$, $\lambda_0 = 1.55$ μm. As in Fig. 8, the solid (red) and dashed (blue) curves refer to the modes of TE- and TM-polarizations, respectively. The shaded area in Figs. 9(a) and 9(b) shows the existence region of the modes in the regime B. The comparison of Figs. 9(a) and 9(b) demonstrates the similar dispersion spectra behavior. Only in the case of TM-polarized EMWs the fundamental mode transforms from the regime A to the regime B with the increase of $L_1$. So in the range $n_{eff}^{TM} < \beta/k_0 < n_{YIG}$ only one TM$_0$-mode of the YIG-layer can propagate, and the PC layer affects the effective refraction index only near the cutoff thickness $L_2$. For the TE-polarization all modes become one-layer modes: in the area $n_{YIG} < \beta/k_0 < n_{eff}^{TE}$ there is a guided mode of the PC-layer, while YIG acts as a cladding of the waveguide structure.

At the inset in Fig. 9 (b) the numbers 1 and 2 denote the TE$_0$- and TM$_0$-modes in the regime B, respectively, which are represented by dotted lines in Fig. 9(c) (curves 1 and 2), while the numbers 3 and 4 refer to the TE$_2$- and TM$_2$-modes in the regime A. Note also, that in contrast to the Fig. 9(a), only TE$_0$-mode is present in the regimes A and B simultaneously.

An important peculiarity of the considered structure is that for a fixed layer thickness the condition $n_{eff}^{TM} < n_{YIG} < n_{eff}^{TE}$ is satisfied, which, in turn, leads to an "inversed" localization of the guided waves: TE-polarized wave propagates in the PC-layer and TM-waves are guided in the YIG-layer.

It should be noted, that the change of ratio $d_1/d_2$ strongly affects the effective refractive index of the PC, and thus the propagation regime of the EMWs.

## 4. Conclusions

We have investigated the dispersion dependencies of the normal electromagnetic waves in the hybrid magneto-optical waveguide structure *SiO$_2$/YIG/(GGG/TiO$_2$)$^N$/vacuum* for transverse magneto-optical configuration. Our calculations demonstrate the pronounced influence of the dielectric PC covering the magneto-optical waveguide on the dispersion characteristics of TE- and TM-modes. It is shown that the presence of the PC leads to an essential distortion of the dispersion curves which increases with growing of the number of the periods of the photonic structure. We also studied the case when period $d$ of the photonic crystal is comparable with the working wavelength $\lambda_0$ of the waveguide and the long-wavelength limit ($d \ll \lambda_0$). It is shown that the relation of the geometrical parameters of the considered structure (especially, the period of the PC) and the working wavelength is crucial factor which affect the behavior of the dispersion curves. We hope that our results will be useful for designing the complex devices (based on magneto-optical waveguides and photonic crystals) for modern integrated photonics. We hope that the investigated above combined structures based on MO waveguides and PCs can open new opportunities for possible applications in photonic devices.

**Acknowledgments**

This research has received funding from the European Union's Horizon 2020 research and innovation programme under the Marie Skłodowska-Curie grant agreement No. 644348 (project "MagIC") (M.K., N.N.D., Yu.S.D., and I.L.L.), COST Action MP1403 "Nanoscale Quantum Optics" (N.N.D., Yu.S.D., and I.L.L.), and also is supported by the grants from the Ministry of Education and Science of Russian Federation: Project No. 14.Z50.31.0015 and No. 3.2202.2014/K (N.N.D., Yu.S.D., I.S.P and D.G.S) and also is supported by Ukrainian State Fund for Fundamental Research under project No. Ф71/73-2016 "Multifunctional Photonic Structures" (I.L.L.).